\documentclass[12pt,tightenlines,eqsecnum,floats,aps,amsmath,amssymb,nofootinbib,superscriptaddress,showpacs]{revtex4}

\usepackage{amsmath,amsthm,amssymb,amsfonts}

\usepackage{dcolumn}
\usepackage{bm}
\usepackage[latin1]{inputenc}
\usepackage[spanish,english]{babel}
\usepackage{amsfonts}
\usepackage{amssymb}
\usepackage{graphicx}
\usepackage{hyperref}

\newcommand{\be}{\begin{equation}}
\newcommand{\ee}{\end{equation}}
\newcommand{\bea}{\begin{eqnarray}}
\newcommand{\eea}{\end{eqnarray}}

\begin{document}
\title{{\bf Revising the observable consequences of slow-roll inflation }}

\author{Iván Agulló}\email{ivan.agullo@uv.es}
\affiliation{ {\footnotesize Physics Department, University of
Wisconsin-Milwaukee, P.O.Box 413, Milwaukee, WI 53201 \  USA}}
\author{José Navarro-Salas}\email{jnavarro@ific.uv.es}
\affiliation{ {\footnotesize Departamento de Física Teórica and
IFIC, Centro Mixto Universidad de Valencia-CSIC.
    Facultad de Física, Universidad de Valencia,
        Burjassot-46100, Valencia, Spain. }}

\author{Gonzalo J. Olmo}\email{olmo@iem.cfmac.csic.es }
\affiliation{ {\footnotesize Instituto de Estructura de la Materia, CSIC, Serrano 121, 28006 Madrid, Spain}}
\affiliation{ {\footnotesize Physics Department, University of
Wisconsin-Milwaukee, P.O.Box 413, Milwaukee, WI 53201 \  USA}}
\affiliation{ {\footnotesize Departamento de Física Teórica and
IFIC, Centro Mixto Universidad de Valencia-CSIC.
    Facultad de Física, Universidad de Valencia,
        Burjassot-46100, Valencia, Spain. }}

\author{Leonard Parker}\email{leonard@uwm.edu}
\affiliation{ {\footnotesize Physics Department, University of
Wisconsin-Milwaukee, P.O.Box 413, Milwaukee, WI 53201 \  USA}}

\date{November 4th, 2009}

\begin{abstract}
We study the generation of primordial perturbations in a (single-field) slow-roll inflationary universe.
In momentum space, these (Gaussian) perturbations are characterized by a zero mean and a non-zero variance $\Delta^2(k, t)$. However, in position space the variance diverges in the ultraviolet. The requirement of a finite variance in position space forces one to regularize $\Delta^2(k, t)$. This can (and should) be achieved by proper renormalization in an expanding universe in a unique way. This affects the predicted scalar and tensorial power spectra (evaluated when the modes acquire classical properties)  for wavelengths that today are at observable scales. As a consequence, the imprint of slow-roll inflation on the CMB anisotropies is significantly altered.  We find a non-trivial  change in the consistency condition that relates the tensor-to-scalar ratio $r$ to the spectral indices. For instance, an exact scale-invariant tensorial power spectrum, $n_t=0$, is now compatible with a non-zero ratio $r\approx 0.12\pm0.06$, which is forbidden by the standard prediction ($r=-8n_t$).
The influence of relic gravitational waves on the CMB may soon come within the range of  planned measurements, offering a non-trivial test of the new predictions.

\end{abstract}


\maketitle

\section{Introduction and summary}

Inflation \cite{inflation} provides a natural solution to the horizon and flatness problems of the hot big-bang cosmology.
A
sufficiently long period of rapid expansion can
explain the large scale homogeneity, isotropy, and flatness of our
visible universe. Inflation also provides a quantitative
explanation \cite {inflation2} to account for the origin of small inhomogeneities in the early universe. These inhomogeneities are responsible for
the structure formation in the universe and for the anisotropies
present in the cosmic microwave background (CMB), which were first
detected by the COBE satellite  and further analyzed by the
WMAP satellite
\cite{WMAP5}.
The potential-energy density of the inflaton field
is assumed to cause the inflationary  accelerated expansion, and the
amplification of its quantum fluctuations and those of the metric are inevitable consequences in an
expanding universe \cite{parker69}. These fluctuations  acquire classical properties in the inflationary period and  provide the initial conditions
for classical cosmological perturbations after the big-bang.
The detection of the effects of primordial tensorial metric fluctuations (gravitational waves) in future high-precision measurements of the
CMB anisotropies
will  serve as a highly non-trivial test of the inflationary paradigm and to constrain specific models. Therefore,
it is particularly important to scrutinize the predictions of inflation for the tensorial and scalar power spectra. In this respect, it was pointed out in \cite{Parker07} (see also \cite{agulloetal08}) that quantum field renormalization significantly modifies  the amplitude of quantum fluctuations, and hence the corresponding power spectra, in de Sitter inflation. The analysis was further improved in \cite{agulloetal09} (see also the essay \cite{essay}) to
 understand how the basic testable predictions of (single-field) slow-roll inflation
 could be affected by quantum field renormalization. In this work we further study this issue, improve the technical analysis, and provide a more complete and robust discussion of how the observable consequences of inflation are altered when quantum field renormalization is taken into account.

Let us briefly summarize the logic of our approach. Let us assume that $\varphi(\vec{x},t)$ represents a perturbation obeying a free field wave-equation on the inflationary background $ds^2= -dt^2+a^2(t)d\vec{x}^2$, where $a(t)$ is a quasi-exponential expansion factor ($a(t) \sim e^{Ht}$ ). At the quantum level, this field is expanded as
\be \label{expansionvarphi}\varphi(\vec{x},t)= \frac{1}{(2\pi)^{3/2}} \int d^3k [\varphi_{k}(t)a_{\vec{k}}e^{i\vec{k}\vec{x}} +  \varphi_{k}^*(t)a^{\dagger}_{\vec{k}}e^{-i\vec{k}\vec{x}}] \ , \ee
where the creation and annihilation operators satisfy the canonical commutation relation $[ a_{\vec{k}}, a^{\dagger}_{\vec{k'}}]=\delta^3 (\vec{k} - \vec{k}')$. The mode functions $\varphi_{k}(t)$ are required to satisfy the adiabatic condition (see, for instance, \cite{parker-toms}).
 The  power spectrum  for this perturbation, $\Delta^2_{\varphi}(k,t)$, is usually defined in terms of the Fourier transform of the variance of the field  \cite{Dodelson, Baumann}
\be \label{variancekk'}\langle \hat \varphi_{\vec{k}}(t) \hat \varphi^{\dagger}_{\vec{k}'}( t)\rangle =  \delta^3 (\vec{k} - \vec{k}')\frac{2\pi^2}{k^3}\Delta^2_{\varphi}(k,t) \ , \ee
where $\hat \varphi_{\vec{k}}(t)\equiv\varphi_{k}(t)a_{\vec{k}}$.
These modes describe a perturbation field characterized, in momentum space, by a zero mean $\langle \hat \varphi_{\vec{k}}(t)\rangle =0$ and the variance (\ref{variancekk'}). The advantage of working in momentum space resides in the fact that different modes fluctuate independently of each other, as explicitly displayed by the presence of the delta function in (\ref{variancekk'}). This way, the quantum field is regarded as an infinite collection of oscillators, each with a different value of $\vec{k}$. In position space the perturbation is also characterized by a zero mean $\langle \varphi(\vec{x},t)\rangle=0$ and a variance (or dispersion) \be \langle \varphi^2(\vec{x},t) \rangle = \frac{1}{(2\pi)^3}\int d^3k d^3k' \langle \hat \varphi_{\vec{k}}(t) \hat \varphi^{\dagger}_{\vec{k}'}(t)\rangle e^{i (\vec{k}-\vec{k'}) \vec{x}}\ , \ee
which, due to spatial homogeneity, turns out to be independent of $\vec{x}$. This variance is formally related to the power spectrum  by
\be \label{tpfunctioncp}\langle \varphi^2(\vec{x},t) \rangle = \int_0^{\infty}\frac{dk}{k} \Delta^2_{\varphi}(k,t) \ . \ee
As is well-known in quantum field theory, the above  expectation value quadratic in the field $\varphi$ is divergent. It suffers from quadratic and logarithmic ultraviolet divergences
\be \label{tpfunctioncpDS}\langle \varphi^2(\vec{x},t) \rangle \sim \frac{1}{4\pi^2}\int_0^{\infty}dk(\frac{k}{a^2} + \frac{\dot a^2}{a^2k} + ...) \ . \ee
 The first term corresponds to the usual contribution from vacuum fluctuations in Minkowski space. This contribution can be eliminated by renormalization as claimed, for instance, in \cite{Lindebook}.  However,  the second term  is also characteristic of vacuum fluctuations in a curved background.
Because the different $k$-modes fluctuate independently of each other, one could be tempted to get rid of this logarithmic ultraviolet divergence by simply eliminating the modes with $k>aH$ and leaving the rest  unaffected (see, for instance, \cite{Lyth-Liddle09}). If one eliminates this divergence using a window function in this way, as is usual for random fields, then one obtains $\Delta^2_{\varphi}(k)\approx H^2/4\pi^2$, where $\Delta^2_{\varphi}(k)$ is defined by the quantity  $\Delta^2_{\varphi}(k,t)$ evaluated a few Hubble times  after the ``horizon crossing time'' $t_k$, since this is the time scale at which the modes behave as classical perturbations.\footnote{The time $t_k$ is defined by $a(t_k)/k=H(t_k)$, where $a(t)$ is the expansion factor and $H=\dot a/a$ is the Hubble rate.} However, one should take into account that the field fluctuations are quantum in nature and, therefore, one should consider the subtle points of quantum field theory (QFT) regarding the ultraviolet divergences.

Even though free quantum field theory is usually regarded as an infinite set of independent harmonic oscillators (one for each k-mode), there are fundamental holistic aspects of QFT that can not be properly understood in terms of independent modes. Renormalization is the hallmark of the holistic aspect of QFT. This is clear in the fact that, although the renormalization schemes in QFT in curved spacetimes are  based on the ultraviolet behavior of the theory,  the infrared sector is also affected by renormalization, leading potentially to observable consequences. This can  be explicitly displayed by considering, for instance, the Casimir effect. The energy density between the two conducting plates obtained by proper renormalization provides the well-known and experimentally tested expression. However, a naive subtraction obtained by introducing a high-frequency cut-off in the integrals in momentum space (i.e., treating the $k$-modes as being independent) produces a quite different result (see the discussion of \cite{hollands-wald}).

Taking this into account, we see that the logarithmic divergence in (\ref{tpfunctioncpDS}) should be dealt with by renormalization and one can not rule out, a priori, the possibility that the treatment of the divergences at very high values of $k$ may produce some impact at lower momenta.
Therefore, we propose that in the standard definitions of the spectrum $\Delta^2_{\varphi}(k, t)$, as given in (\ref{variancekk'}) and (\ref{tpfunctioncp}), one should replace the unrenormalized $\langle \varphi^2(\vec{x},t) \rangle$ by the renormalized variance, $\langle \varphi^2(\vec{x},t) \rangle_{ren}$. Writing $\tilde\Delta^2_{\varphi}(k,t)$ for the spectrum defined in this way,  the definition in (\ref{tpfunctioncp}) (and similarly in (\ref{variancekk'})) is replaced by the corresponding renormalized expression

\be \label{varren}\langle \varphi^2(\vec{x},t) \rangle_{ren}   = \int_0^{\infty}\frac{dk}{k} \tilde\Delta^2_{\varphi}(k,t) \ . \ee
 This way, the physical variance $\langle \varphi^2(\vec{x},t) \rangle_{ren}$ remains a well-defined quantity, in the same way as one could obtain a finite expression for the expectation values of the quantum stress-energy tensor. To complete the physical consistency of this approach, it would be desirable to define a unique expression for the necessary subtractions required to produce a consistent $\langle \varphi^2(\vec{x},t)\rangle_{ren}$. Since the power spectrum is defined in momentum space, the natural scheme is renormalization in momentum space, so we define
 \be \label{varrendef}\langle \varphi^2(\vec{x},t)\rangle_{ren}   = \frac{4\pi}{(2\pi)^3}\int_0^{\infty}k^2dk(|\varphi_{k}(t)|^2 - C_{k}(t)) \ , \ee
where $C_{k}(t)$ represents the expected counterterms. As we will see, adiabatic renormalization \cite{parker-fulling, parker-toms, birrel-davies}, which works by subtracting a set of counterterms ``mode-by-mode'', provides a natural expression for the counterterms encoded in $C_{k}(t)$. Moreover, the DeWitt-Schwinger renormalization, originally defined in position space (see, for instance, \cite{parker-toms, birrel-davies}), can be nicely translated to momentum space \cite{Bunch-Parker}, thus providing another answer for $C_{k}(t)$. When these two schemes are applied to the field perturbations arising from inflation, the resulting expressions for  $C_{k}(t)$ coincide, thus defining a unique expression for  the spectrum $\tilde\Delta^2_{\varphi}(k,t)$.

The holistic nature of QFT is then explicitly realized through (\ref{varren}). Although the counterterms are fully determined by the ultraviolet behavior of the modes,  the long wavelength sector, and hence the new $\tilde\Delta^2_{\varphi}(k,t)$, is significantly affected by the  subtractions.
In the slow-roll scenario, when $H$ slowly decreases with time, the effects of renormalization have a non-trivial impact on $\tilde\Delta^2_{\varphi}(k,t)$ when this quantity is evaluated a few Hubble times after the time of horizon crossing $t_k$. For instance, for the tensorial modes we obtain
\be  \tilde\Delta^2_{}(k) \sim \left(\frac{H}{2\pi}\right)^2 \epsilon  \ , \ee
where $\epsilon$ is the usual slow-roll parameter. A similar expression is obtained for the scalar perturbations (involving now the slow-roll $\epsilon$ and $\eta$ parameters), with the corresponding changes in the tensor-to-scalar ratio, the spectral indices and the consistency relation. The new
predictions remain in agreement with observation for the
simplest forms of inflation ($\phi^2$ and $\phi^4$ potentials).
It is worth pointing out that the new consistency condition for single-field inflation is predicted  to be \be \label{relcons}  r=4(1-n_s-n_t)+\frac{4n'_t}{n_t^2-2n'_t} \left(1-n_s-\sqrt{2 n'_t+(1-n_s)^2-n_t^2} \right) \ , \ee
instead of the standard prediction $r=-8n_t$. The tensor-to-scalar ratio is now related with the spectral indices $n_t$, $(1-n_s)$, and also $n'_t\equiv dn_t/d\ln{k}$. This modification has far reaching consequences. For instance, since the observations from the 5-year WMAP \cite{WMAP5}(with BAO+SN) strongly suggest that $(1-n_s) \approx 0.030 {\pm 0.015}$ (with $r<0.22$), expression (\ref{relcons}) allows for an exact scale invariant tensorial power spectrum, $n_t=0$, while being compatible with a non-zero ratio $r\approx 0.12\pm0.06$. We will comment on this further later on.

The paper is organized as follows. In section II we briefly review the standard ways to derive the expressions for the tensor and scalar power spectra of single-field slow-roll inflation.
In section III we work out the new definition of the power spectra and give the technical details involving the required renormalization in momentum space. In section IV we provide, as a consequence of the new power spectra, the corresponding expression for the tensor-to-scalar ratio $r$, the tensorial and scalar spectral indices, and the slow-roll parameters. This leads, in particular, to a
change in the consistency condition that relates the
tensor-to-scalar amplitude ratio to the spectral indices. In section V we summarize our results and conclusions. We use natural units $\hbar=1=c$.

\section{Spectrum of fluctuations from inflation}

We will now proceed to briefly review the standard predictions for the power spectra in single-field slow-roll inflation.  In obtaining these predictions,  quantum field renormalization in the curved spacetime of the expanding universe is not taken into account.

\subsection{Tensorial spectrum}
Let us focus on the production of relic gravitational waves by considering fluctuating tensorial modes $h_{ij}(\vec{x},t)$ in an expanding, spatially flat universe
\be ds^2= -dt^2 + a^2(t)(\delta_{ij} + h_{ij})dx^idx^j \ . \ee
The wave equation obeyed by these modes comes from the linearized Einstein equations and is given by
\be -a^2 \ddot h_{ij} -3a\dot a \dot h_{ij} + \nabla^2 h_{ij} =0  \ . \ee
Expanding the fluctuating fields $h_{ij}$ in plane wave modes $h_k(t) e_{ij}e^{i\vec{k}\vec{x}}$,
where $e_{ij}$ is a constant polarization tensor obeying the conditions $e_{ij}=e_{ji}, e_{ii}=0$ and $k_ie_{ij}=0$,
we obtain the equation
\be \label{waveq}
\ddot{h}_k + 3H \dot{h}_k +\frac{k^2}{a^2} h_k
=0 \ , \ee
with $k\equiv |\vec{k}|$.
The conditions for the polarization tensor imply that the
perturbation field $h_{ij}$ can be decomposed into two polarization
states described by a couple of  massless scalar fields $h_{+,
\times}(\vec{x},t)$, both obeying the wave equation (\ref{waveq}) \cite{Lifshitz}
 (see also, for instance, \cite{books}; from now on, we omit the
subindex $+$ or $\times$).
On scales larger than the Hubble radius, the damping term $3H \dot{h}_k$ dominates. However,
on scales smaller than the Hubble radius it is the spatial gradient term that dominates over the damping term, thus leading to the conventional flat-space oscillatory behavior of modes. To constrain the form of the modes defining the quantization,  it is natural to impose the adiabatic asymptotic condition \cite{parker69,parker-toms} for large $k$
\be \label{asymptotic}\frac{h_{k}(t)}{\sqrt{16 \pi G}}  \sim (2(2\pi)^3 w(t)a^3(t))^{-1/2}e^{-i\int^t w(t')dt'} \ , \ee with $w(t)=k/a(t)$, where the factor $\sqrt{16 \pi G}$ ($G$ is the Newton constant) has to be introduced to get a canonically normalized variable. This condition does not uniquely fix the form of the modes. For instance, in an exact de Sitter background ($a(t)=e^{Ht}$, with $H$ a stric constant) one can  invoke
 de Sitter invariance  to uniquely determine the modes \cite{bunch-davies}, and one obtains  \be \label{modesh}
h_{k}(t) = \sqrt{\frac{16\pi G}{2
(2\pi)^3k^3}} ( H
-ike^{-Ht})e^{i(kH^{-1}e^{-Ht})} \ . \ee
 These modes oscillate until the physical wave length reaches the Hubble length $H^{-1}$. The amplitude of the modes at this time, usually called the ``horizon exit'' time $t_k$, defined by $k/a(t_k)=H$, is then $|h_{{k}}|^2 = \frac{ 2GH^2}{\pi^2k^3}$.  A few Hubble times after horizon exit, the modes get frozen as classical perturbations \cite{books, Dodelson, Lyth-Liddle09} (see also \cite{KieferPointerStates2007}) with constant amplitude $|h_{{k}}|^2 = \frac{ GH^2}{\pi^2k^3}$.   The freezing amplitude is
usually codified through the unrenormalized quantity $\Delta_h^2(k)= 4\pi k^3 |h_{{k}}|^2$.
Taking
into account the two polarizations,   one easily gets the standard
scale free tensorial power spectrum  $P_t(k) \equiv 4\Delta_h^2(k)=
\frac{8}{M_P^2} \left (\frac{H}{2\pi}\right)^2$, where $M_P=1/\sqrt{8\pi G} $ is the reduced Planck mass in natural units.
The appearance of such frozen fluctuations converts the modes into classical perturbations with wavelengths that are still stretched by the rapid expansion to reach astronomical scales. This is essentially what happens in the inflationary era.

\subsubsection{Slow-roll inflation}

 To take into account that inflation lasts for a finite period of time, one usually considers the so-called slow roll scenario \cite{Dodelson, books, Lyth-Liddle09}. The homogeneous part of the inflaton field $\phi_0(t)$ rolls slowly down its potential $V(\phi)$ towards a minimum.
Both $\phi_0$ and $H\approx \sqrt{\frac{8\pi G}{3}V(\phi_0)}$ are changing very gradually and this change is parameterized by the slow-roll parameters $\epsilon, \eta$, where $\epsilon=-\dot{H}/H^2$, and $\eta-\epsilon = \ddot{\phi_0}/(H\dot{\phi}_0)$. In the slow-roll approximation, defined when the parameters are small $\epsilon, |\eta|\ll1$, one can relate them to the derivatives of the inflaton potential $\epsilon=(M_P^2/2)(V'/V)^2, \eta = M_P^2(V''/V)$.
To generate an approximate form for the modes, it is convenient to introduce the  conformal time variable $\tau\equiv\int dt/a(t)$. In terms of $\tau$ the wave equation turns out to be of the form
\be \frac{d^2 h_k}{d\tau^2}+ 2Ha\frac{d h_k}{d\tau}+ k^2h_k=0 \ , \ee
and taking into account that in the slow-roll approximation
\be \label{srapp}(1-\epsilon)\tau= -\frac{1}{aH} \ , \ee
 we get
\be \frac{d^2 h_k}{d\tau^2} -\frac{2(1+\epsilon)}{\tau}\frac{d h_k}{d\tau}+ k^2h_k=0 \ . \ee
Within this approximation, and treating now the parameter $\epsilon$ as a constant, one can exactly solve the above equation as follows
\be \label{amplten} h_{k}(t) = (-16\pi G\tau
\pi/4(2\pi)^3a^2)^{1/2}[E(k)H^{(1)}_{\nu}(-k\tau)+ F(k) H^{(2)}_{\nu}(-k\tau)]\ , \ee
where the index of the Bessel function is  $\nu=3/2 + \epsilon$, and the complex coefficients $E(k)$ and $F(k)$ obey the normalization requirement
\be |E(k)|^2 - |F(k)|^2 =1 \ . \ee
The adiabaticity condition (\ref{asymptotic}) implies that
\be \lim_{k\to \infty} E(k) =1 \ \ \ \  \lim_{k\to \infty} F(k) =0  \ . \ee
The simplest way to choose $E(k)$ and $F(k)$ would be to require that, for $\epsilon \to 0$, we recover the exact de Sitter form of the modes. This would mean that $E(k)=1$ and $F(k)=0$ for every value of $k$.

\subsubsection{Infrared divergences}

The choice $E(k)=1$ and $F(k)=0$ is, however, an idealized situation which assumes that inflation started at an infinite time in the past.
A consequence of this assumption is that
the two-point function
\be \label{2pt}\langle 0|h(x)h(x')|0\rangle =
\int d^3k e^{i\vec{k}(\vec{x}-\vec{x'})}h_k(\tau) h_k^*(\tau') \ee
is ill-defined due to an infrared divergence. This is so because in the limit $k \to 0$ the integrand in (\ref{2pt}) behaves as
\be \frac{dk}{k}\frac{1}{k^{2\epsilon}} \ , \ee and $\epsilon > 0$.
For general $E(k)$ and $F(k)$ we have, instead,
\be \frac{dk}{k}\frac{|E(k)-F(k)|^2}{k^{2\epsilon}} \ . \ee
The infrared divergence is avoided if $|E(k)-F(k)|^2 \to 0$, as $k \to 0$. This  happens naturally if one assumes that inflation started smoothly at some early, but finite, time (see, for instance, \cite{Glenz-Parker09}).  Assuming that the initial vacuum is well-defined, and since the dynamical evolution cannot generate infrared divergences \cite{ford-parker, fulling-wald}, one should get $E(k) \to F(k)$, when $k \to 0$ (as explicitly obtained in \cite{Glenz-Parker09}), and a finite contribution to the two-point function in the infrared end.

The scale of this infrared behavior of the functions $E(k)$ and $F(k)$ is given by the Hubble radius at the beginning of inflation. Only those wavelengths that were already outside the inflationary Hubble radius
\be k/a(t) \ll H(t) \ee
when inflation began can see this infrared behavior. However, only the wavelengths that crossed the Hubble radius
\be  k/a(t_k)=H(t_k) \ee
at $50-60$ $e$-folds before the end of the inflationary epoch have cosmological observable consequences today.
For these wavelengths the functions $E(k)$ and $F(k)$ behave as in the Bunch-Davies vacuum ($E(k) \sim 1, F(k) \sim 0$). Despite having no direct observable consequences, the behavior of $E(k)$ and $F(k)$ for very large wavelengths is important for the consistency of the theoretical framework because it avoids an infrared divergence in the two-point function.

\subsubsection{Nearly scale-invariant  spectrum}

The Hubble exit time $t_k$ can be rewritten, in terms of the conformal time, as
\be -k\tau(t_k)= (1+\epsilon) \ . \ee Therefore,
 at the Hubble exit time $t_k$ the amplitude of the modes $h_{k}$ given in (\ref{amplten}) takes the value $|h_k|^2 = \frac{2GH^2(t_k)}{\pi^2 k^3}$, and then $\Delta_h^2(k)= \frac{8GH^2(t_k)}{\pi}$. This provides a nearly scale-invariant spectrum of fluctuations. The $k$ dependence of $H^2(t_k)$
 \be \frac{d \ln H(t_k)}{d\ln k}= -\epsilon \ee
 leads to \be \Delta_h^2(k) =  \Delta_h^2(k_0)\left(\frac{k}{k_0}\right)^{-2\epsilon} \ , \ee where $k_0$ is a pivot scale. One defines the tensorial spectral index $n_t$ as the exponent in the above expression. So \be n_t=-2\epsilon (t_k)\ . \ee
In the latter formula  we have explicitly considered that the constant parameter $\epsilon$ in the previous calculation is given by its value around the Hubble exit for the $k$ mode.

  The same conclusion can be achieved by evaluating the power spectrum a few $e$-foldings  after crossing the Hubble radius. One can then consider the approximation
$-k\tau \ll 1$, which implies
\be |H^{(1)}_{\nu}(-k\tau)|^2 \sim \frac{2}{\pi}(-k\tau)^{-2\nu} \ . \ee
Furthermore, from the slow-roll equation (\ref{srapp}) we obtain
\be  \label{referencetau0}a=  a(\tau_0)\left(\frac{\tau}{\tau_0}\right)^{\frac{1}{\epsilon -1}}\ , \ee
where $\tau_0$ is an arbitrary reference time.
Therefore, we can write
\be \label{hh} | h_k(
t)|^2 \sim  \frac{16\pi G
}{4(2\pi)^3}\frac{(-\pi \tau)}{a^2} \frac{2}{\pi}(-k\tau)^{-2\nu} \sim \frac{G
}{\pi^2 k a^2(\tau_0)}(-k\tau_0)^{-2-2\epsilon} \ , \ee
which explicitly shows the dependence on $k^{-2\epsilon}$ of $\Delta_h^2(k)\propto k^3| h_k(
t)|^2$.
Now, taking for convenience the reference time $\tau_0$ in (\ref{referencetau0}) as $\tau_0=\tau(t_k)$ and using  that
$a(t_k)=k/H(t_k)$ and $\tau(t_k) = -\frac{1}{(1-\epsilon)k}$,
one can find immediately that
\be \label{h}| h_k(
t)|^2 \sim \frac{G
}{\pi^2 k^3} H^2(t_k) \left(\frac{1}{1-\epsilon}\right)^{-2-2\epsilon}\approx  \frac{G
}{\pi^2 k^3} H^2(t_k) \ , \ee
Taking into account the two polarizations one finally gets the same unrenormalized  power spectrum  $P_t(k)\equiv 4\Delta_h^2(k)=16\pi k^3 | h_k|^2$ (up to a factor $1/2$)
\be P_t(k)= \frac{8}{M_P^2}\left (\frac{H(t_k)}{2\pi} \right )^2
\ . \ee

\subsection{Scalar spectrum}

We now proceed to reproduce the standard results for the unrenormalized power spectrum of scalar perturbations. We first consider, for simplicity, the inflaton field in an unperturbed background and compute the $\Delta^2_{\varphi}$ evaluated at $t_k$. In order to evaluate $\Delta^2_{\varphi}$ at later times (a few $e$-folds after $t_k$), we must improve the calculation by taking into account the fluctuation of the background metric using the spatially-flat slicing. We also show how to reproduce this result using a more rigorous approach in terms of gauge invariant quantities.

Consider the inflaton field $\phi(t,\vec x )$ made out of a homogeneous part $\phi_0(t)$ and a small fluctuating part $\delta
\phi(\vec{x},t)$. The fluctuation $\delta
\phi(\vec{x},t)$ satisfies, in the approximation of considering an unperturbed background spacetime, the wave equation
\be \label{scalarwe}\ddot{\delta \phi} + 3H
 \dot{\delta \phi} -a^{-2}\nabla^2 \delta \phi
+ V''(\phi_0)\delta \phi =0 \ . \ee
Due to the slow-rolling of $\phi_0(t)$ from the hill potential,
the  term $V''$ is very small, which allows one to estimate the amplitude of quantum
fluctuations $\Delta_{\phi}^2$ in a way similar to $\Delta_h^2$.
Inflaton fluctuations translate into curvature perturbations, which
constitute the ``seeds'' for structure formation and are
characterized by their scalar power spectrum $P_{\cal R}(k) =
\left (H/\dot{\phi_0}\right )^2\Delta_{\phi}^2 $. Evaluating $\Delta_{\phi}^2 $ and $\left (H/\dot{\phi_0}\right )^2$ at the time of horizon exit $t_k$
\cite{books, Dodelson}  one gets $\Delta_{\phi}^2 = H^2(t_k)/2\pi^2$ and $H^2(t_k)/\dot{\phi_0}^2(t_k)= \frac{1}{2M^2_P\epsilon(t_k)}$. Therefore,
$P_{\cal R}(k)\sim \frac{1}{M^2_P\epsilon(t_k)}\left (\frac{H(t_k)}{2\pi} \right )^2$.

One can improve this approximation by taking into account the effect of the metric perturbation in (\ref{scalarwe}). To properly define the wave equation for $\delta
\phi(\vec{x},t)$ we have to specify a slicing. Choosing for convenience the spatially-flat slicing, the equation for the field perturbation
 $\delta
\phi(\vec{x},t)$   is (see, for instance, \cite{Lyth-Liddle09})
\be \label{scalarweLL}\ddot{\delta \phi} + 3H
 \dot{\delta \phi} -a^{-2}\nabla^2 \delta \phi
+ V''(\phi_0)\delta \phi +\frac{1}{a^3}\frac{d}{dt}\left(\frac{2a^3\dot H}{H}\right) \delta \phi =0 \ . \ee
In the slow-roll approximation, the corresponding Fourier components of $\delta \phi$ obey
 \be \label{scalarwe3}\ddot{\delta \phi_k} + 3H
 \dot{\delta \phi_k} +a^{-2}k^2 \delta \phi_k
+ H^2(3\eta -6\epsilon)\delta \phi_k =0 \ . \ee
To deal with this equation, one usually considers that the $\epsilon$ and $\eta$ parameters are constant, while $H=\dot a/a$ remains a time-dependent function. So we have
\be \label{scalarwe4}\ddot{\delta \phi_k} + 3\frac{\dot a}{a}
 \dot{\delta \phi_k} +a^{-2}k^2 \delta \phi_k
+ \frac{\dot a^2}{a^2}(3\eta -6\epsilon)\delta \phi_k =0 \ . \ee
With this assumption one can exactly solve the equation in terms of the  conformal time $\tau$.
The form of the modes $\delta \phi_k$ is the same as for the tensorial ones, up to the coefficient $\sqrt{16\pi G}$,
\be \delta \phi_{k}(t) = (-\tau
\pi/4(2\pi)^3a^2)^{1/2}H^{(1)}_{\mu}(-k\tau) \ , \ee
 where the index of the Bessel function  is now $\mu=3/2 + 3\epsilon -\eta$. Evaluating this perturbation a few Hubble times after the horizon crossing time, one obtains the standard result \be  P_{\cal{R}}(k)=\left (H/\dot{\phi_0}\right )^2 4\pi k^3 |\delta \phi_{{k}}|^2= \frac{1}{2M_P^2\epsilon(t_k)}\left
(\frac{H(t_k)}{2\pi}\right)^2 \ , \ee
where the constant parameter $\epsilon$ is given at the Hubble radius time scale $t_k$. Due to the $k$-dependence of $H(t_k)$ and $\epsilon(t_k)$, one also gets a nearly scale-free spectrum
\be P_{\cal R}(k) = P_{\cal R}(k_0)\left(\frac{k}{k_0}\right)^{n_s-1} \ee
where the so-called scalar spectral index $n_s$ is found to be \be n_s= 1 -6\epsilon +2\eta \ . \ee

To complete our presentation, we will explain how the above result can be recovered   without specifying any particular slicing and threading of spacetime. These scalar perturbations are
commonly studied through the gauge-invariant quantity $\cal{R}$ (the
comoving curvature perturbation)
\be {\cal{R}}= \Psi+\frac{H}{\dot\phi_0}\delta \phi \ , \ee
where $\Psi$ is the curvature perturbation ($R^{(3)}=4\nabla^2\Psi /a^2$) of  the spatial metric $g_{ij}=a^2[(1-2\Psi)\delta_{ij}+2\partial_{ij}E]$. In momentum space, $\cal{R}$ obeys  the
 equation \cite{Mukhanov86,Weinberg2008}\be \frac{d^2 {\cal{R}}_k}{d{\tau}^2} +
\frac{2}{z}\frac{dz}{d\tau}\frac{d {\cal{R}}_k}{d{\tau}} +
k^2{\cal{R}}_k =0 \ , \ee where $z\equiv a\dot{\phi}_0/H$. In the slow roll approximation $z^{-1}dz/d\tau= a H(1+2\epsilon-\eta)$ and hence
\be \label{z} z(\tau)= z(\tau_0) \left(\frac{\tau}{\tau_0}\right)^{-1-3\epsilon +\eta} \ , \ee
where $\tau_0$ is again a reference instant of time.
Furthermore, the above wave equation simplifies to
 \be \label{MSsr}\frac{d^2 {\cal{R}}_k}{d{\tau}^2} -
\frac{2(1+3\epsilon - \eta)}{\tau}\frac{d {\cal{R}}_k}{d{\tau}} +
k^2{\cal{R}}_k =0 \ . \ee
The solution obeying the adiabatic condition (and the de Sitter symmetry for $H$ constant) is
\be {\cal{R}}_k( t) = (-\pi \tau /4(2\pi)^3 z^2)^{1/2}H^{(1)}_{\mu}(-\tau k) \ , \ee  where $\mu= 3/2 + 3\epsilon -\eta$. Evaluating ${\cal{R}}_k( t)$ at $t=t_k$ one obtains the power spectrum estimated above in terms of the fluctuations of the inflaton field.
We can also evaluate the amplitude $|{\cal{R}}_k|^2$ a few Hubble times after $t_k$. Using $\tau(t_k)$ as reference time in (\ref{z}),
it is easy to get ($P_{\cal
R}(k)=\Delta_{\cal R}^2(k)=4\pi k^3 |{\cal{R}}_k|^2$) \be \label{psR}P_{\cal
R}(k) = \frac{1}{2M_P^2\epsilon(t_k)}\left
(\frac{H(t_k)}{2\pi}\right)^2 \ .\ee

\subsection{Tensor-to-scalar amplitude ratio and spectral indices}

We have seen that the concrete value of the amplitude of the power spectra depends on the time at which they are evaluated. In the literature, one find different times that can be parameterized by $\lambda$, according to the condition $k/a(t)=\lambda H$. The power spectra  are then modified by the factor $(1+\lambda^2)$ (see for instance \cite{padmanabhan05}), which goes to unit exponentially fast. This ambiguity, however, is irrelevant in the evaluation of physical observables such as the tensor-to-scalar ratio $r$ and the spectral indices. In fact, irrespective of the evaluation time ($t_k$ or a few $e$-foldings after it), the ratio
\be r= \frac{P_t(k)}{P_{\cal
R}(k)} \ee
gives a constant quantity
\be r=16\epsilon(t_k) \ . \ee
The other two physical observables are neither affected by the ambiguity in the evaluation time of the power spectra. The spectral indices are unambiguously given by
\be   n_t = -2\epsilon(t_k) \ , \ee and \be   1-n_s= 6\epsilon(t_k) - 2\eta(t_k) \ . \ee
From the above formulas one infers a necessary relation between measurable quantities
\be r=-8n_t \ , \ee which, as claimed many times in the literature, should be verified by any single-field slow-roll inflationary model irrespective of the form of the potential.
For this reason, the future experimental checking of this condition is usually regarded as an important test of the simplest forms of inflation.

This concludes our review of the standard derivation of the unrenormalized predictions of single-field slow-roll inflation.

\section{The role of ultraviolet divergences:  renormalization of the spectrum of fluctuations}

In this section we reexamine the derivation of the power spectra taking into account the effects of renormalization.

\subsection{Ultraviolet divergences and momentum-space renormalization}\label{subsectionA}

It is easy to see that $\Delta_h^2(k,t)$ gives the formal contribution, per $d\ln k$, to the variance of the gravitational wave fields $h_{+,
\times}$
 \be \label{varianceh}
\langle h^2 \rangle =\int_0^{\infty}k^2dk \int d\Omega |h_k|^2=
\int_0^{\infty}\frac{dk}{k}\Delta^2_h (k,t) \ ,
\ee
since, as already defined, $\Delta^2_h (k,t)\equiv 4\pi k^3 |h_k|^2$. Although, as noted previously, the small $k$ behavior of the functions $E(k)$ and $F(k)$ cures the potential infrared divergence in the above integral, the large $k$ behavior of the modes makes the  integral divergent
\be \label{divergentintegral}\langle h^2 \rangle =
\int_0^{\infty}\frac{dk}{k} \frac{16\pi G k^3}{4\pi^2a^3}\left [
\frac{a}{k}[1 +\frac{(2+3\epsilon)}{2k^2\tau^2}] + ... \right] \ ,
\ee
It is a common view  to
bypass this point by regarding $h(\vec{x},t)$ as a classical random field. One then introduces a window function $W(kR)$ in the integral to smooth out the field on a certain scale $R$ and to remove the Fourier modes with $k^{-1} < R$.
However, as explained in the introduction, it is our view to regard the variance as a basic physical object and treat $h$ and $\cal{R}$ (or $\delta \phi$ in the flat-slicing gauge) as a proper quantum field. Renormalization is then the natural solution to eliminate the ultraviolet divergences and keep the variance in position space finite and well-defined.
Since the physically relevant quantity (power spectrum) is expressed
in momentum space, the natural renormalization scheme to apply is
the so-called adiabatic subtraction \cite{Parker07}, as it
renormalizes the theory in momentum space. Adiabatic renormalization
\cite{parker-fulling, parker-toms, birrel-davies} removes the
divergences present in the formal expression (\ref{varianceh}) by
subtracting counterterms mode by mode in the integral (\ref{varianceh}) \be
\label{variancehren} \langle h^2 \rangle_{ren} =
\int_0^{\infty}\frac{dk}{k}\left[4\pi k^3|h_{\vec{k}}|^2 - \frac{16\pi G
k^3}{4\pi^2 a^3}(w_k^{-1}+ (W_k^{-1})^{(2)})\right]\ ,\ee with $w_k=k/a(t)$. The
subtraction of the first term $(16\pi G k^3/4\pi^2 a^3 w_k)$
cancels the typical  flat space vacuum fluctuations, which are responsible for the quadratic divergence in the integral (\ref{divergentintegral}).
The additional term, proportional to $(W_k^{-1})^{(2)}$  and which  involves
$\dot{a}^2$ and $\ddot{a}$, is necessary to properly perform the
renormalization in an expanding universe. It cancels the logarithmic divergence in (\ref{divergentintegral}).

However, one can legitimately ask if the momentum-space counterterms are uniquely fixed. In other words, can a different renormalization scheme, as the DeWitt-Schwinger subtraction prescription in momentum space \cite{Bunch-Parker}, lead to a different expression for the counterterms?
Before going further in our analysis, let us briefly summarize the main steps of momentum-space renormalization (for an extensive exposition see \cite{parker-toms}).
\subsection{Momentum-space renormalization}
\subsubsection{Adiabatic renormalization}

Let us consider a generic free scalar field $\varphi$ in our spatially flat cosmological metric obeying the field equation
\be \label{wead}\ddot{\varphi} + 3H
 \dot{\varphi} -a^{-2}\nabla^2 \varphi
+ (m^2 + \xi R)\varphi  =0 \ , \ee
where $R$ is the four-dimensional scalar curvature
\be R=6\left[\left(\frac{\dot a}{a}\right)^2+ \frac{\ddot a}{a}\right] \ , \ee
and $\xi$ is the curvature coupling. We include this term to be general and for illustrative purposes. The Fourier components of $\varphi$ obey the equation
\be \label{scalarwevarphik}\ddot{ \varphi_k} + 3\frac{\dot a}{a}
 \dot{ \varphi_k} +a^{-2}k^2  \varphi_k
+ \left[6\left(\frac{\dot a^2}{a^2}+\frac{\ddot a}{a}\right)\xi+ m^2\right] \varphi_k =0 \ . \ee
Note that the role of the dimensionless parameter $\xi$ here is similar to the role of the $\epsilon$ and $\eta$ parameters in the wave equation (\ref{scalarwe4}) for the perturbation.

Let us assume that the form of the normalized modes is $(2Va^3(t))^{-1/2}e^{i\vec{k}\vec{x}}\bar{\varphi}_k(t)$,
where $V\equiv L^3$ is the volume of a box of coordinate length $L$. The continuous limit is obtained by replacing $V$ by $(2\pi)^3$ and $\sum_{\vec{k}}$ by $ \int d^3\vec{k}$. The function $\bar{\varphi}_k(t)$ obeys the equation
\be \label{WKBeq}\frac{d^2}{dt^2}\bar{\varphi}_k + \Omega_k^2 \bar{\varphi}_k=0 \ , \ee
where
\be \Omega_k^2 = w_k^2 + \sigma \ , \ee
with $w_k=\sqrt{k^2/a^2 + m^2}$ and
\be \label{sigmaxim}\sigma= \left(6\xi-\frac{3}{4}\right)\left(\frac{\dot a}{a}\right)^2+ \left(6\xi -\frac{3}{2}\right)\frac{\ddot a}{a} \ . \ee
 The function $\bar{\varphi}_k(t)$ is assumed to obey also the adiabatic condition $\bar{\varphi}_k(t)\sim w_k^{-1/2}e^{-i\int w_k(t')dt'}$ in the large $k$ regime. This condition does not uniquely  determine the form of $\bar{\varphi}_k(t)$ (different solutions lead to different sets of modes and, therefore, to different vacuum states) but it uniquely determines an asymptotic expansion for all possible solutions. This expansion is characterized by
 \be \bar{\varphi}_k(t) \sim W_k^{-1/2}e^{-i\int W_k(t')dt'} \ , \label{eq:Asymptotic}\ee
 with a recursive expansion
 \be W_k= w_k + w_k^{(2)}+ w_k^{(4)} + ...  \ee
where
\be w_k^{(2)}\equiv  \left(\frac{1}{2}w_k^{-1/2}\frac{d^2}{dt^2} w_k^{-1/2} + \frac{1}{2}w_k^{-1}\sigma\right) \ . \ee
Similar expressions for  $w_k^{(4)}$ and all the other higher-order terms can be found. Each order in the (adiabatic) expansion is characterized by the number of time derivatives of $a(t)$ appearing in a term. While $w_k$ is of zero order, $w_k^{(2)}$ is of second order, as one can trivially verify by simple counting. Note that $\sigma$ is of second adiabatic order. One should have in mind that the adiabatic expansion is an asymptotic series and does not converge in general. This is why the form of the modes is not uniquely defined by the adiabatic condition, thus allowing to have different solutions and hence different vacuum states.\\

The evaluation of the variance of the field $\varphi$ as a sum in modes leads to a divergent expression. In the continuous limit it is given by
\be \langle \varphi^2\rangle = (4\pi^2a^3)^{-1}\int_0^{\infty}dkk^2|\bar{\varphi}_k|^2 \ . \ee
As  is evident from the asymptotic expansion (\ref{eq:Asymptotic}), one necessarily encounters ultraviolet divergences in the above quantity. Generically one encounters quadratic and logarithmic divergences. In the adiabatic renormalization, the physically relevant finite expression is obtained from the formal one by subtracting mode by mode each term in the adiabatic expansion of the integrand that contains at least one ultraviolet divergent part for arbitrary values of the parameters ($m$ and $\xi$) of the theory \cite{parker-toms}. Applying this to the particular case of the variance of the field $\varphi$ one gets
 \be \langle \varphi^2\rangle_{ren} = (4\pi^2a^3)^{-1}\int_0^{\infty}dkk^2\left(|\bar{\varphi}_k|^2 - w_k^{-1} - (W_k^{-1})^{(2)}\right)\ , \ee
where $w_k^{-1}$ and $(W_k^{-1})^{(2)}$ are the zeroth and second order terms, respectively, in the adiabatic expansion of $W_k^{-1}$.
Generically the second adiabatic counterterm $(W^{-1})^{(2)}$ is given  by
\be (W_k^{-1})^{(2)} = -\frac{1}{w_k^2}\left[\frac{1}{2}w_k^{-1/2}\frac{d^2}{dt^2}w_k^{-1/2} + \frac{1}{2}w_k^{-1}\sigma\right] \ . \ee
Note that, in the flat space limit $\bar{\varphi}_k$ goes to the usual Minkowski form of the modes, $(W_k^{-1})^{(2)}$ goes to zero, and $\langle \varphi^2\rangle_{ren}$ gives a vanishing result, in agreement with normal ordering. In a dynamical universe, $(W_k^{-1})^{(2)}$ is generically nonzero and the renormalization  produces a non-trivial, well-defined result for the physical variance.

For its physical relevance in our reevaluation of the inflationary power spectra,  it is specially interesting to  analyze the massless limit of the above result. When $m \to 0$, the subtraction counterterms
take the form
\be \label{count} (4\pi^2a^3)^{-1}k^2\left(w_k^{-1} + (W_k^{-1})^{(2)}\right)=(4\pi^2a^3)^{-1}k^2 \left(\frac{a}{k}+ (1-6\xi)\frac{a^3}{2k^3}\left[\left(\frac{\dot a}{a}\right)^2+ \frac{\ddot a}{a}\right]\right) \ . \ee

\subsubsection{ Comparison with the momentum-space representation of the DeWitt-Schwinger subtraction terms}

Using a local momentum representation in a normal neighborhood admitting Riemann normal coordinates, one can express in momentum space the DeWitt-Schwinger proper time representation of the Green functions \cite{Bunch-Parker} (see also \cite{birrel-davies, parker-toms}). Up to second adiabatic order, we have
\be G^{(2)}_{DS}(P, Q) = -i\frac{|g(P)|^{1/4}}{(2\pi)^4}\int d^4k e^{iky}[\frac{1}{(k^2+m^2)}+ \frac{(\frac{1}{6}-\xi)R}{(k^2+m^2)^2}]  \ , \ee
where $g(P)$ is the determinant of the metric at the point $P$ in the so-called Riemann normal coordinates $y^{\mu}$, $ky\equiv -k_0y^0+ \vec{k}\vec{y}$, and the contour of integration in the $k_0$ plane is assumed to be the usual contour defining the particular two-point function. The point $Q$ has been taken as the reference point for constructing the Riemann coordinates.
These coordinates  are constructed by considering the unique geodesic that joins the reference point with an arbitrary point $P$ in a normal neighborhood of $Q$. The Riemann coordinates $y^{\mu}$ of $P$ are given by
\be y^{\mu}= \lambda \xi^{\mu} \ , \ee
where $\lambda$ is the value at $P$ of an affine parameter of the geodesic joining $Q$, at $\lambda=0$, to $P$. The vector $\xi^{\mu}$ is the tangent to the geodesic at the point $Q$
\be \xi^{\nu}=\frac{dx^{\mu}}{d\lambda}\Big|_{Q} \ . \ee
Although in these coordinates the form of the geodesic equations is trivial, the form of the metric at $Q$ is not necessarily  Minkowskian. When we impose, additionally, that $ds^2|_{Q}= -(dy^0)^2 + (dy^1)^2+ (dy^2)^2+ (dy^3)^2$, we have then the so-called Riemann normal coordinates.
In our spatially flat universe $ds^2= -dt^2 + a^2(t)d\vec{x}^2$ and for our problem, we take the initial and final points ($Q\equiv (t_0, \vec{x}'), P\equiv (t_0, \vec{x})$) at the constant time hypersurface $t=t_0$. We then have $y^0=0$ and
\be \vec{y} = a(t_0)(\vec{x} -\vec{x}')\ . \ee Moreover $g(P)=g(Q) =1$.

The explicit form of $G^{(2)}_{DS}(P, Q) $ in our cosmological scenario reduces, after performing the $k_0$ integration, to
\be G^{(2)}_{DS}(\vec{x}, t_0;\vec{x}', t_0)= I_1 - (\frac{1}{6}-\xi)RI_2 \ , \ee
where $R$ is the scalar curvature and
\be I_1= \frac{1}{2(2\pi)^3}\int d^3\vec{k} \frac{e^{i\vec{k}(\vec{y}-\vec{y}')}}{w_{ka}} \ , \ee
\be I_2= \frac{1}{2(2\pi)^3}\int d^3\vec{k} \frac{e^{i\vec{k}(\vec{y}-\vec{y}')}}{2w_{ka}^3} \ , \ee
with $w^2_{ka}=k^2
+ m^2$. Finally, performing the change of variables $\vec{k} \to \vec{k}/a(t_0)$ we get
\be G^{(2)}_{DS}(\vec{x}, t_0;\vec{x}', t_0)=\frac{1}{2(2\pi)^3a^3}\int d^3\vec{k} \left[\frac{1}{w_{k}}+ \frac{(\frac{1}{6}-\xi)R}{2w^3_{k}}\right]e^{i\vec{k}(\vec{x}-\vec{x}')} \ . \ee
In the DeWitt-Schwinger point-splitting framework, the renormalization of the variance $\langle \varphi^2(\vec{x},t_0)\rangle$ proceeds by subtracting  $G^{(2)}_{DS}(\vec{x}, t_0;\vec{x}', t_0)$ to the two-point function and taking the coincidence-point limit $\vec{x}' \to \vec{x}$
\be \langle \varphi^2(\vec{x},t_0)\rangle_{ren} = \lim_{\vec{x}' \to \vec{x}}[\langle \varphi(\vec{x},t_0)\varphi(\vec{x}', t_0)\rangle - G^{(2)}_{DS}(\vec{x}, t_0;\vec{x}', t_0)]\ . \ee
In the massless limit, $m \to 0$, the integrand of the momentum-space representation of $G^{(2)}_{DS}(\vec{x}, t_0;\vec{x}', t_0)$ is
\bea G^{(2)}_{DS}(\vec{x}, t_0;\vec{x}', t_0)&=&\frac{1}{2(2\pi)^3a^3}\int d^3\vec{k} \left(\frac{a}{k}+ (1-6\xi)\frac{a^3}{2k^3}\left[\left(\frac{\dot a}{a}\right)^2+ \frac{\ddot a}{a}\right]\right)e^{i\vec{k}(\vec{x}-\vec{x}')}\nonumber \\ &=& \frac{1}{4\pi^2 a^3}\int_0^{\infty}dkk^2\left(\frac{a}{k}+ (1-6\xi)\frac{a^3}{2k^3}\left[\left(\frac{\dot a}{a}\right)^2+ \frac{\ddot a}{a}\right]\right)\frac{\sin k |\vec{x}-\vec{x}'|}{k|\vec{x}-\vec{x}'|} \ . \eea
Taking $\vec{x}\to\vec{x}'$, the integrand of the above expression coincides exactly with the momentum-space counterterms obtained previously via the adiabatic renormalization. Therefore, the question ending section \ref{subsectionA} is answered. The adiabatic subtraction coincides with the DeWitt-Schwinger subtraction prescription in momentum space in the massless limit, irrespective of the value of the dimensionless parameter $\xi$.

\subsection{Tensorial spectrum}
Let us apply the above scheme to the tensorial fluctuations of the metric.
To determine the counterterm $(W_k^{-1})^{(2)}$  one should  rewrite the  wave equation (\ref{waveq}) in the form (\ref{WKBeq}).
This can be easily obtained by performing the change $\bar{h}_k= a^{3/2} h_k$. One obtains $w_k=k/a$ (i.e., $m=0$) and a second order adiabatic term $\sigma$ of the form (\ref{sigmaxim}) with $\xi=0$
\be \sigma= -\frac{3}{4}\left(\frac{\dot a}{a}\right)^2-\frac{3}{2}\frac{\ddot a}{a} \ . \ee
A straightforward calculation gives
\be (W_k^{-1})^{(2)}=\frac{\dot{a}^2}{2a^2w_k^3} +\frac{\ddot{a}}{2aw_k^3} \ . \ee
To perform the explicit computation it is useful to take into account the slow-roll relations
\begin{eqnarray}
{\frac{da}{d\tau}}&=&\frac{(1+2\epsilon )}{(H \tau^2)}\\
{\frac{d^2a}{d\tau^2}}&=&-\frac{(2+5\epsilon )}{(H \tau^3)}
\end{eqnarray}
These derivatives are related with the usual dotted derivatives as follows:
\begin{eqnarray}
\dot a &=& a^{-1}\frac{da}{d\tau}\\
\ddot a &=& -a^{-3}\left(\frac{da}{d\tau}\right)^2 + a^{-2}\frac{d^2a}{d\tau^2}
\end{eqnarray}
The final expression for zeroth and second order adiabatic  counterterms is then\footnote{In a similar way, the  DeWitt-Schwinger subtraction terms in momentum-space lead to the same result. Note in passing that neither of the proposed (adiabatic) counterterms defined in \cite{durrer09}  for the tensorial and scalar power spectra seem to agree with those obtained here by adiabatic regularization.}
\be w_k^{-1}+(W_k^{-1})^{(2)}=\frac{a}{k}\left[1+\frac{(2+3\epsilon)}{2k^2\tau^2}\right] \ . \ee
The subtraction of these counterterms produces the expression (assuming that $E(k)=1$ and $F(k)=0$ for every $k$)\footnote{Note that the second adiabatic counterterm introduces an  infrared divergence in the continuum limit. For $k \to 0$ the integrand in $\langle h^2 \rangle_{ren}$ is proportional to
$dk/k$. This logarithmic divergence disappears in the finite box formulation and has no impact on the observable power spectrum, in contrast to the ultraviolet divergences. Furthermore, the calculation of the renormalized stress-energy tensor $\langle T_{\mu\nu}\rangle$, and hence the evolution of the universe, is insensitive to the infrared cutoff $L$ when $L \to \infty$.}
\bea \langle h^2 \rangle_{ren} &=& \int_0^{\infty}\frac{dk}{k} \frac{16\pi G
k^3(-\tau \pi)}{4\pi^2 2a^2} \left(|H^{(1)}_{\nu}(-k\tau)|^2
- \frac{2}{\pi(-k\tau)}\left[1+\frac{(2+3\epsilon)}{2k^2\tau^2}\right]\right)\ . \eea
Therefore, the renormalized expression for $\Delta_h^2(k,t)$ is\
\be \tilde\Delta_h^2(k,t)= \frac{16\pi G
k^3(-\tau \pi)}{4\pi^2 2a^2} \left(|H^{(1)}_{\nu}(-k\tau)|^2
- \frac{2}{\pi(-k\tau)}\left[1+\frac{(2+3\epsilon)}{2k^2\tau^2}\right]\right) \ . \ee
Note that the asymptotic behavior for large $k$ of the Hankel function is
\be  |H^{(1)}_{\nu}(-k\tau)|^2 \sim \frac{2}{\pi (-k\tau)}\left[1
+\frac{(4{\nu}^2 -1)}{8(-k\tau)^2}\right] + ...\ . \ee
The first two terms for $\nu=3/2 + \epsilon$ are exactly the adiabatic counterterms subtracted above.

It is useful to remark that, in adiabatic renormalization (or, equivalently, in the DeWitt-Schwinger subtraction algorithm), the number of terms that should be subtracted is determined by the degree of divergence of the expectation value under consideration. In our case, this degree of divergence is of second-order. Therefore, one must stop necessarily at second order if the purpose is to make finite the expectation value $\langle h^2 \rangle$. The relevant subtractions are carried out mode-by-mode and result in a sum-over-modes that has no ultra-violet divergence. In addition, as we have already stressed, it is important to point out that it is necessary to subtract the relevant adiabatic counterterms for all modes, including those with large wavelength, even if the expansion of the universe is not slow. Note that  the reason for the name``adiabatic'' regularization is a result of the fact that the adiabatic (or arbitrarily slow) limit is used to identify the necessary subtractions. Therefore, it is not correct to argue that the adiabatic renormalization terms are invalid when the expansion is rapid, i.e., in the non-adiabatic regime.

\subsection{Scalar spectrum}

We will work out the (renormalized) spectrum of scalar perturbations using the gauge-invariant formalism. To convert equation (\ref{MSsr}) into one of the form (\ref{WKBeq}) we have to perform the change $\bar{\cal{R}}_k=a^{1/2}z{\cal{R}}_k$.
We then obtain that $w_k=k/a$, and that the second order adiabatic function $\sigma$ is given by
\be \label{sigmascalar}\sigma=  -\frac{3}{4}\left(\frac{\dot a}{a}\right)^2-\frac{3}{2}\frac{\ddot a}{a} + (3\eta -6\epsilon)\left(\frac{\dot a}{a}\right)^2\ . \ee
Note that this $\sigma$ is different from that of (\ref{sigmaxim}).
The reason for this can be easily seen by looking at the wave equation
 in the flat-slicing gauge (\ref{scalarwe4}). That equation does not
have a constant massive term, and the second adiabatic order term is
$H^2(3\eta -6\epsilon)$, which is just the last term in
(\ref{sigmascalar}).
This implies that the scalar perturbation (the inflaton in  the
flat-slicing gauge or $\cal{R}$ in the gauge-invariant
approach)  should be regarded as a massless field.\footnote{This was first taken into account in the calculation of the renormalized power spectra in \cite{agulloetal09, Talk-MG12} in terms of the  gauge-invariant quantity $\cal{R}$. This quantity was also quantized in \cite{Sugiyama}, but there the $V''$ term in (\ref{scalarweLL}) was treated as zeroth adiabatic order, unlike the present treatment, in the evaluation of the counterterms.} The corresponding second order counterterm is
thus of the form
\be (W_k^{-1})^{(2)}=\frac{\dot{a}^2}{2a^2w_k^3} +\frac{\ddot{a}}{2aw_k^3} -\frac{1}{2w_k^3}(3\eta -6\epsilon)(\frac{\dot a}{a})^2 \ . \ee
Therefore, the final expression for the counterterms is
\be w_k^{-1} + (W_k^{-1})^{(2)}=\frac{a}{k}\left[1 + \frac{(2+3(3\epsilon-\eta))}{2(-k\tau)^2}\right] \ . \ee

Proceeding in a parallel way as for the  tensorial case, we get a renormalized value for $\langle {\cal{R}}^2 \rangle$
\be  \langle {\cal{R}}^2 \rangle_{ren} =
\int_0^{\infty}\frac{dk}{k}\left[4\pi k^3|{\cal{R}}_{k}|^2 - \frac{
k^3}{4\pi^2 z^2a}(w_k^{-1}+ (W_k^{-1})^{(2)})\right]\ . \ee
Therefore,
\be \label{varianceRren} \langle {\cal{R}}^2 \rangle_{ren} = \int_0^{\infty}\frac{dk}{k} 4\pi k^3
\frac{-\pi \tau}{ 4(2\pi)^3 z^2}\left(|H^{(1)}_{\mu}(-\tau k)|^2 - \frac{2}{\pi (-k\tau)}\left[1 + \frac{(2+3(3\epsilon-\eta))}{2(-k\tau)^2}\right]\right)\ ,  \ee
where the index $\mu$ is given by $\mu=3/2 +3\epsilon -\eta$.
The renormalized expression for $\Delta_{\cal{R}}^2(k,t)$ is
\be \label{renorDeltaR} \tilde\Delta_{\cal{R}}^2(k,t)=
\frac{4\pi k^3(-\pi \tau)}{ 4(2\pi)^3 z^2}\left(|H^{(1)}_{\mu}(-\tau k)|^2 - \frac{2}{\pi (-k\tau)}\left[1 + \frac{(2+3(3\epsilon-\eta))}{2(-k\tau)^2}\right]\right)\ .  \ee
Note finally that, as expected, the same result is obtained working in the flat-slicing gauge. One then obtains
\be \tilde\Delta_\phi^2(k,t)= \frac{
k^3(-\tau \pi)}{4\pi^2 2a^2} \left(|H^{(1)}_{\mu}(-\tau k)|^2 - \frac{2}{\pi (-k\tau)}\left[1 + \frac{(2+3(3\epsilon-\eta))}{2(-k\tau)^2}\right]\right) \ , \ee
where $\mu=3/2 + 3\epsilon -\eta$. Taking into account that $\tilde\Delta_{\cal{R}}^2(k, t)= (a^2/z^2) \tilde\Delta_\phi^2(k,t)$  one recovers the expression (\ref{renorDeltaR})  for $\tilde\Delta_{\cal{R}}^2(k, t)$.

\section{Testable consequences}

In this section we work out new expressions for the observable magnitudes that follow from the renormalized power spectra of the previous section.

\subsection{Tensor-to-scalar ratio}

Let us now consider the tensor-to-scalar ratio
\be r\equiv 4\frac{\tilde\Delta_h^2(k,t)}{\tilde\Delta_{\cal{R}}^2(k,t)}= 4\frac{16\pi G
k^3(-\tau \pi)}{4\pi^2 2a^2} \frac{ 4(2\pi)^3 z^2}{4\pi k^3(-\pi \tau)}\frac{\left(|H^{(1)}_{\nu}(-\tau k)|^2 - \frac{2}{\pi (-k\tau)}\left[1 + \frac{(2+3\epsilon)}{2(-k\tau)^2}\right]\right)}{\left(|H^{(1)}_{\mu}(-\tau k)|^2 - \frac{2}{\pi (-k\tau)}\left[1 + \frac{(2+3(3\epsilon-\eta))}{2(-k\tau)^2}\right]\right)}\ee
Since at the Hubble exit time $t_k$ we have $z^2(t_k)=2M_P^2\epsilon(t_k)a^2(t_k)$, it follows that
\be 4\frac{16\pi G
k^3(-\tau \pi)}{4\pi^2 2a^2} \frac{ 4(2\pi)^3 z^2}{4\pi k^3(-\pi \tau)}\Big|_{t=t_k}=16 \epsilon(t_k) \ . \ee
We also find that
\be \left.\left(|H^{(1)}_{\nu}(-\tau k)|^2 - \frac{2}{\pi (-k\tau)}\left[1 + \frac{(2+3\epsilon)}{2(-k\tau)^2}\right]\right)\right|_{t=t_k} = \frac{2}{\pi}\alpha \epsilon (t_k) \ , \ee
where $\alpha$ is a numerical constant of order unity, $\alpha \approx 0.904$.
A similar estimation is obtained for the factor coming from the scalar spectrum (up to the substitution $\nu \to \mu$, i.e., $\epsilon \to 3\epsilon - \eta$)
\be \left.\left(|H^{(1)}_{\mu}(-\tau k)|^2 - \frac{2}{\pi (-k\tau)}\left[1 + \frac{(2+3(3\epsilon-\eta))}{2(-k\tau)^2}\right]\right)\right|_{t=t_k} =\frac{2}{\pi}\alpha (3\epsilon (t_k)-\eta(t_k)) \ . \ee
Therefore, the tensor-to-scalar ratio $r$, evaluated at the Hubble exit time, is
\be \label{rrenormalized}r= 16 \epsilon(t_k)\frac{\epsilon (t_k)}{3\epsilon (t_k)-\eta(t_k)} \ . \ee

An important comment is now in order. To obtain this result we have evaluated the power spectra at the Hubble exit time. Nevertheless, as we will show in the next paragraph, this estimate does not depend critically on the precise time at which the counterterms are evaluated. During slow-roll inflation the  counterterms decay as $|\tau|^{2 \epsilon}$ (tensorial), $|\tau|^{2(3 \epsilon -\eta)}$ (scalar), in the late-time limit as $|\tau|\to 0$. This means that they decay very slowly and are in fact constant in exact de Sitter inflation. After reheating, the counterterms decay more rapidly, as is obvious from (\ref{count}). One would then recover the standard prediction $r=16\epsilon(t_k)$. However, since the modes acquire classical properties soon after exiting the Hubble sphere, the relevant time to evaluate these magnitudes falls in the interval between $t_k$ and a few $e$-foldings after it.

Let us analyze in detail the time dependence of $\tilde\Delta_h^2(k,t)$, $\tilde\Delta_{\cal{R}}^2(k,t)$, and $r$  in terms of the number $n$ of e-folds after crossing the Hubble radius. Since $-k\tau(t_k)=1+\epsilon$, we can write $-k\tau$
as
\be \label{eq:ktau}  -k\tau=  (1+\epsilon)\frac{\tau}{\tau(t_k)}=(1+\epsilon)(\frac{a(t_k)}{a})^{\frac{1}{(1+\epsilon)}}= (1+\epsilon)e^{-\frac{n}{(1+\epsilon)}} \ . \ee
The expression for $\tilde\Delta_h^2(k,t)$ when $-k\tau \ll 1$ leads to
 \bea \tilde\Delta_h^2(k,t)&\approx& \frac{16\pi G
k^3(-\tau \pi)}{4\pi^2 2a^2} \frac{2}{\pi}(-k\tau)^{-2\nu}[1
- (-k\tau)^{2\nu}
\frac{(2+3\epsilon)}{2(-k\tau)^3}]] \nonumber \\
&\approx& \frac{2}{M_P^2}\left(\frac{H(t_k)}{2\pi}\right)^2[1
- (-k\tau)^{2\nu}
\frac{(2+3\epsilon)}{2(-k\tau)^3}]
\ , \eea
and using (\ref{eq:ktau}), we find
\be \tilde\Delta_h^2(k,n)\approx \frac{2}{M_P^2}\left (\frac{H(t_k)}{2\pi} \right )^2[1
-\frac{(2+3\epsilon)}{2}e^{-2\epsilon n}]\ . \label{eq:Dh-n}\ee
Assuming that we are just a few $e$-folds  after the Hubble exit but before the end of inflation, i.e., $n>1$ but $n\epsilon \ll 1$, so that $e^{-2\epsilon n}\approx 1 -2\epsilon n$, we obtain\footnote{We note that if one evaluates the power spectra at the end of the slow-roll era (where $n\epsilon \sim 1$) the contribution of the counterterms is still significant. However, we find it more natural to evaluate the spectra soon after Hubble exit, when the modes have already acquired classical properties.}
\be \tilde\Delta_h^2(k,n)\approx \frac{2}{M_P^2}\left (\frac{H(t_k)}{2\pi} \right )^2 \epsilon(t_k) (2n-3/2)\ . \label{eq:Dh-n2}\ee
A similar estimation can be obtained for the scalar power spectrum
\be \tilde\Delta_{\cal{R}}^2(k,n)\approx \frac{1}{2M_P^2\epsilon(t_k)}\left
(\frac{H(t_k)}{2\pi}\right)^2 (3\epsilon(t_k) - \eta(t_k))(2n-3/2)\ . \label{eq:DR-n}\ee
Note that  the parameter $n$ enters in the power spectra parameterizing the (unknown) time at which the modes exhibit classical behavior.
However, since both tensorial and scalar spectra have the same dependence on $n$, the tensor-to-scalar ratio is not sensitive to the unknown parameter $n$, in the same way as it is insensitive to the scale of inflation $H(t_k)$, and it is essentially given by (\ref{rrenormalized}). The same conclusion can be drawn if one estimates the time derivative of  $r$. One then obtains that

\be \frac{dr}{dn}\approx O(\epsilon^2) \ . \ee

This means that the renormalized value for $r$ is changing very slowly,
in agreement with the previous evaluation. In other words, the value of $r$ evaluated at the Hubble radius crossing time  remains nearly constant during this period of inflation. Obviously it is not strictly constant, as in the computation without renormalization, but its change is slow.\footnote{To be precise, $d(\ln r)/dn=2 \epsilon-\eta$. Note that for the exponential potential model $V(\phi)\propto \exp [-(\phi/M_P)\sqrt{2/p}]$ we have $\eta=2\epsilon$. Then the new predictions coincide, by accident, with the standard ones. So in this case $\dot r =0$.} We regard this result as a signal of the robustness of renormalization in determining the spectra of inflationary perturbations.

\subsection{Spectral indices}

The above calculations have another important consequence: the spectral indices  remain unchanged when they are evaluated at $t_k$ or a few Hubble times after it. By definition, we have
\be n_t\equiv \frac{d\ln P_t}{d\ln k}= 2\frac{d\ln H(t_k)}{d\ln k}+\frac{d\ln \epsilon(t_k)}{d\ln k} \ , \ee
which, according to (\ref{eq:Dh-n2}), turns out to be independent of $n$. The result is
\be \label{nt} n_t= 2(\epsilon-\eta) \ , \ee
where $\epsilon$ and $\eta$ are evaluated at $t_k$.
For the scalar index we have
\be \label{ns}n_s-1\equiv  \frac{d\ln P_{\cal{R}}}{d\ln k}= 2\frac{d \ln H(t_k)}{d\ln k} - \frac{d \ln \epsilon(t_k)}{d\ln k}+\frac{d \ln (3\epsilon(t_k)- \eta(t_k))}{d\ln k}  \ . \ee
Explicit evaluation gives
\be n_s-1=-6 \epsilon+2 \eta+\frac{ (12 \epsilon^2-8\epsilon \eta+\xi) }{3 \epsilon-\eta}\ee
where $\xi$ is another slow roll parameter \cite{Dodelson}: $\xi\equiv M_P^4(V'V'''/V^2)$. This parameter can be reexpressed in terms of $\epsilon, \eta$ and the running of the tensorial index $n'_t\equiv dn_t/d\ln k$ as follows
\be \label{ntt} n'_t= 8\epsilon (\epsilon - \eta)+2\xi \ . \ee

\subsection{Consistency condition}

The formulas (\ref{rrenormalized}, \ref{nt}-\ref{ntt}) provide an algebraic relation between the tensor-to-scalar ratio and the spectral indices, $r=r(n_t, n_s, n'_t)$, that takes the following form
\be \label{eq:r} r=4(1-n_s-n_t)+\frac{4n'_t}{n_t^2-2n'_t} \left(1-n_s-\sqrt{2 n'_t+(1-n_s)^2-n_t^2} \right) \ . \ee
Positivity of the argument of the square root in this expression imposes the following constraint
\be (1-n_s)^2-n_t^2 + 2 n'_t\geq 0 \ . \label{eq:ineq}\ee

The consistency relation (\ref{eq:r}), contrasts with that obtained without invoking renormalization,  namely, $r=-8n_t$. Since $r$ must be positive, according to the standard prediction one would expect $n_t$ to be negative. This restriction, however, does not follow either from (\ref{eq:r}) nor from (\ref{eq:ineq}). Even more, from (\ref{eq:ineq}) it is easy to see that our prediction allows for $n_t\geq 0$, in sharp contrast with the standard prediction, which necessarily requires $n_t=-2\epsilon <0$. We also remark that, according to the standard derivation, the running of $n_t$ is fully determined by the values of $n_s$ and $n_t$, since then one finds $n'_t=-n_t(1-n_s + n_t)$. On the contrary, the manipulations that led to (\ref{eq:r}) indicate that $n'_t$ is now an independent quantity that needs to be measured in order to check the new consistency relation (\ref{eq:r}). This aspect could make more challenging the experimental verification of the consistency condition of (single-field) slow roll inflation. \\

 We stress again  that the new consistency relation allows for a null tensorial tilt $n_t =0$ while being compatible with a non-zero ratio $r\approx 4(1-n_s)$. Since the observations from the 5-year WMAP \cite{WMAP5}(with BAO+SN) strongly suggest that $(1-n_s) \approx 0.030 {\pm 0.015}$ (with $r<0.22$)\footnote{Marginalized over all other parameters of a flat $\Lambda$CDM model.} it follows that, for the case of an exact scale invariant tensorial power spectrum, $n_t=0$, our consistency relation  leads to $r\approx 0.12\pm0.06$. These possibilities  may  soon come within the measurement range of  forthcoming  CMB polarization experiments such as the PLANCK satellite, the QUIJOTE CMB experiment, SPIDER, Polar BEAR, EBEX, BICEP, SPUD, and the future CMBPol mission \cite{experiments}.

\subsection{Comparison between the standard and the new predictions}

We can compare the differences between the standard and the new  predictions in terms of the representative set of chaotic potential models $V(\phi)= \lambda M_P^{(4-p)} \phi^p$.
 The standard prediction is
\be \label{rstandard}
r=\frac{4p}{N}  \ \ , \ \ 1-
n_s= \frac{(p+2)}{2N} \ \ , \ \ n_t= -\frac{p}{2 N}    \ \ , \ \ n'_t=\frac{p}{2 N^2}     \ , \ee
 where $N\equiv \ln{a_{end}/a(t_k)}$ is the
number of $e$-folds of inflation between the horizon crossing time $t_k$ of cosmological wavelengths and the end of inflation.
If we invoke renormalization we get, instead,
 \be \label{r}
r=\frac{4 p^2}{(p+2)N}  \ \ , \ \ 1-
n_s= \frac{p}{2N} \ \ , \ \ n_t=  \frac{(2-p)}{2 N}  \ \ , \ \ n'_t= \frac{(2-p)}{2 N^2}    \ . \ee
Note that now the quadratic potential is characterized by having an exact scale-invariant behavior for the tensorial spectrum.

\subsection{Comparison with WMAP data}

Although a definite test of the new predictions, like the above proposed consistency condition, requires more accurate data on tensor fluctuations (so far we have only upper limits on $r$) we can compare the new predictions with the standard ones on the basis of the five year WMAP data. We can contrast the predictions (\ref{r}) for $r$ and $n_s$, for the representative values $p=2$ and $p=4$,  with the WMAP
5-year data (see Fig. 5 of Ref. \cite{WMAP5}). We find that  both models are compatible with the experimental data for the
reasonable range of $N$ between $50$ and $60$. This is in
sharp contrast with the prediction of the standard approach,
 where the monomial potential with $p=4$ is
excluded convincingly.

\section{Conclusions and final comments}

Inflationary cosmology predicts that, due to quantum
effects, small density perturbations are generated in the very early
universe with a nearly ``scale-free'' spectrum. The detection and
analysis of anisotropies in the cosmic microwave background has
 confirmed this prediction. Moreover,  inflation also
predicts the creation of primordial gravitational waves, which still
remain undetectable. Forthcoming high-precision measurements of the
cosmic microwave background \cite{experiments} may measure effects of relic gravitational
waves, and this will be crucial to test the inflationary paradigm and strongly
constrain inflationary models. Therefore, it is of crucial importance to scrutinize, from all points of view, the quantitative
predictions of inflation.

In this work we have pointed out that, if quantum
field renormalization is taken into account, the predictions of (single-field)
slow-roll inflation for both the scalar and tensorial power spectra
 change significantly. In our physical context, renormalization is naturally implemented in a mode-by-mode subtraction scheme and uniquely defines the momentum-space counterterms. These counterterms are evaluated in the period when the perturbations acquire classical properties and this leads to testable predictions  that differ  significantly  from the standard ones.
 Because of the question of the
underlying quantum nature of the gravitational field, there are several ways one could think
about the process by which the dispersion spectrum of the inflaton field influences the
spectrum of the scalar perturbations of the metric. One way is to think of the gravitational field
as making a measurement of each mode of the inflaton fluctuation field within a few Hubble times
of its exit from the Hubble  sphere. If this measurement of the inflaton dispersion spectrum is similar to standard measurements in quantum mechanics, then it should measure the renormalized value of
the inflaton spectrum at the time when the measurement is carried out, which we take to be the time when the perturbations acquire significant classical properties. The time $t$ at which this occurs affects both the scalar and tensorial power spectra, but for a large range of values of $t$, the effect of renormalization remains significant for
wavelengths that today are at observable scales. We have shown that if the power spectra are evaluated $n$ $e$-folds after the Hubble radius exit time, where $n\ll 1/\epsilon$, then the observable parameters $r$ and $n_s, n_t, n'_t$ are insensitive to the value of $n$. If the power spectra were evaluated at a time $t$ well after the end of inflation, one would recover the standard predictions. With the present understanding of the
nonlinear aspects of quantum gravity, it is difficult to reach a definitive answer regarding the value
of $n$, so the fact that we find observable
differences offers a deep way to experimentally probe this question.\\

\noindent { \bf Acknowledgements.} This
work has been partially supported by the spanish grant FIS2008-06078-C03-02. I.A. and L.P.  have been partly
supported by NSF grants PHY-0071044 and PHY-0503366 and by
a UWM RGI grant.  G.O.
thanks MICINN for a JdC contract and the ``Jos\'e Castillejo'' program for funding a stay at the University of Wisconsin-Milwaukee.
We thank D. Lyth for interesting correspondence.
We also  thank R. Durrer, G. Marozzi, and M. Rinaldi for a private communication
about their views on our results in \cite{agulloetal09}.
\\


\begin{thebibliography}{99}

\bibitem{inflation} Guth A.,  {\it Phys. Rev.} D{\bf 23}, 347 (1981).
%
Starobinsky A.A., {\it Phys. Lett.} B{\bf 91}, 99 (1980).
%
Linde, A.D., {\it Phys. Lett.} B{\bf 108}, 389 (1982); {\it Phys. Lett.} B{\bf 129}, 177 (1983).
%
Albrecht A. and  Steinhardt P. J., {\it Phys. Rev. Lett.} {\bf 48},1220 (1982).
%
Sato, K.,  {\it Mon. Not. Roy. Astron. Soc.} {\bf 195}, 467 (1981).
\bibitem{inflation2} Mukhanov V.F. and Chibisov G.V., {\it JETP Letters} {\bf33}, 532(1981).
%
Hawking, S. W., {\it Phys. Lett.} B{\bf 115}, 295 (1982).
%
Guth A. and Pi, S.-Y., {\it Phys. Rev. Lett.} {\bf 49}, 1110 (1982).
%
Starobinsky, A. A., {\it Phys. Lett.} B{\bf 117}, 175 (1982).
%
Bardeen, J.M., Steinhardt, P.J. and Turner, M.S., {\it Phys. Rev.}D{\bf 28}, 679 (1983).
\bibitem{WMAP5} Komatsu, E. {\it et al.}, {\it J. Suppl. Ser.} {\bf 180}, 330 (2009).
\bibitem{parker69} Parker L., {\it Phys.Rev.Lett.} {\bf 21} 562 (1968); {\it Phys. Rev.} {\bf 183}, 1057(1969); Parker L., {\it The creation of particles in an expanding universe}, Ph.D. thesis, Harvard University (1966).

\bibitem{Parker07} Parker L.,  {\ Amplitude of perturbations from inflation}, hep-th/0702216.
\bibitem{agulloetal08} Agull\'o I., Navarro-Salas J.,  Olmo G.J. and  Parker L., {\it Phys. Rev. Lett.} {\bf 101}, 171301 (2008).

\bibitem{agulloetal09} Agull\'o I., Navarro-Salas J.,  Olmo G.J. and  Parker L., {\it Phys. Rev. Lett.} {\bf 103}, 061301 (2009).
\bibitem{essay} Agull\'o I., Navarro-Salas J.,  Olmo G.J. and  Parker L.,  {\it Gen. Rel. Grav.} {\bf 41}, 2301 (2009). 

\bibitem{parker-toms} Parker L. and Toms D.J., {\it Quantum field theory in curved spacetime: quantized fields
and gravity}, Cambridge University Press, (2009).


\bibitem{Dodelson}
Dodelson S., \textit{Modern Cosmology}, Academic Press, (2003). Liddle A. R., and Lyth D.H., \textit{Cosmological inflation and large-scale structure}, Cambridge University Press, (2000).
\bibitem{Baumann} Baumann D., {\it TASI Lectures on inflation}, arXiv:0907.5424. Baumann D., et al. {it CMPPolMission Concept Study}, arXiv:0811.3919
  \bibitem{Lindebook} Linde A.,{\it Particle physics and inflationary cosmology}, CRC Press (1990). Padmanabhan T., {\it Structure formation in the universe}, Cambridge University Press (1993).
\bibitem{Lyth-Liddle09} Lyth D.H. and Liddle A.R. \textit{The primordial density perturbation}, Cambridge University Press, (2009).
\bibitem{hollands-wald}Hollands C. and  Wald R., {\it Quantum field theory is not merely quantum mechanics applied to the low energy effective degrees of freedom}, arXiv:gr-qc/0405082v1.

 \bibitem{parker-fulling} Parker L. and  Fulling S.A., {\it Phys. Rev. D} {\bf 9} 341 (1974). Fulling S.A. and Parker L., {\it Ann. Phys. (N.Y.)} {\bf 87} 176 (1974). Fulling S.A., Parker L. and Hu B.L., {\it Phys. Rev. D} {\bf 10} 3905 (1974). See also LP's Ph.D. thesis given in \cite{parker69}.

\bibitem{birrel-davies} Birrel N.D. and Davies P.C.W., {\it Quantum fields in curved space}, Cambridge University Press, (1982).
\bibitem{Bunch-Parker} Bunch T.S. and Parker L. {\it Phys. Rev. D} {\bf 20} 2499 (1979).

\bibitem{Lifshitz} Lifshitz E.M., {\it Zh. Eksp. Teor. Fiz.} {\bf 16}, 587 (1946).

\bibitem{books} Kolb E.W. and Turner M.S. {\it The early universe}, Westview Press (1990).
 \bibitem{bunch-davies} Bunch T.S. and Davies P.C.W., {\it Proc. Roy. Soc.} {\bf A}360, 117 (1978).
\bibitem{KieferPointerStates2007} Kiefer C., Lohmar I., Polarski D., Starobinsky A. A., Class. Quant. Grav. {\bf 24} 1699 (2007).



\bibitem{Glenz-Parker09}  Glenz M. and  Parker L., {\it Phys. Rev. D} {\bf 80} 063534 (2009). Glenz M., {\it Topics in Inflationary Cosmology and Astrophysics}, Ph.D.Thesis, arXiv:0905.2641 .

 \bibitem{ford-parker}Ford L.H. and Parker L., {\it Phys. Rev. D} {\bf 16} 245 (1977).
\bibitem{fulling-wald} Fulling, S.A., Sweeny, M. and  Wald, R.M. {\it Commun. Math. Phys.} {\bf 63} 257 (1978).


 \bibitem{Mukhanov86} Mukhanov V.S., {\it JETP Lett.} {\bf 41}, 493 (1986). Sasaki S. {\it Prog. Theor. Phys.} {\bf 76}, 1036 (1986)
\bibitem{Weinberg2008} Weinberg S., \textit{Cosmology}, Oxford University Press, (2008).


\bibitem{padmanabhan05}Sriramkumar L., Padmanabhan T., {\it Phys.Rev.D} {\bf 71} 103512 (2005).



\bibitem{durrer09}Durrer R., Marozzi G., Rinaldi M., {\it Phys.Rev.D} {\bf 80} 065024 (2009)
\bibitem{Talk-MG12} Agull\'o I., Navarro-Salas J.,  Olmo G.J. and  Parker L., {\it Inflation, quantum field renormalization, and CMB anisotropies.} Talk given by J. Navarro-Salas in the 12th Marcel Grosmann Meeting; Paris, 12-18, July (2009). To appear in the proceedings of MG12.

\bibitem{Sugiyama} Sugiyama N., {\em The Amplitude of Perturbation during Inflation and the Initial Conditions for the Evolution of Radiation and Matter in the Universe}, M.S. thesis, University of Wisconsin-Milwaukee (2008).

\bibitem{experiments}
 \url{http://www.rssd.esa.int/index.php?project=planck};\\
 \url{http://www.iac.es/project/cmb/quijote/};\\
 \url{http://cmb.phys.cwru.edu/ruhl_lab/spider.html};\\
 \url{http://bolo.berkeley.edu/polarbear/};\\
 \url{http://groups.physics.umn.edu/cosmology/ebex};\\
 \url{http://cmbpol.uchicago.edu/ }.


\end{thebibliography}
\end{document}